\documentclass[journal=jpcafh,manuscript=article]{achemso}
\usepackage[version=3]{mhchem} 
\usepackage{graphicx}
\usepackage{float}
\usepackage{cancel} 
\usepackage{amsmath}
\usepackage{amssymb}
\usepackage{easybmat}
\usepackage{booktabs}
\usepackage{siunitx}
\usepackage{multirow}
\usepackage{algorithm}
\usepackage{algpseudocode}
\usepackage{tikz}
\usepackage{algpseudocode}
\usetikzlibrary{positioning}
\usepackage{bm}
\usetikzlibrary{shapes,arrows,arrows,positioning,fit}
\usepackage{xurl}
\usepackage{hyperref}
\usepackage{xcolor}
\usepackage{soul}
\usepackage{svg}
\usepackage{makecell}

\usepackage{subcaption}
\usepackage{placeins}

\usepackage{listings}
\usepackage{subcaption}

\def\icm{cm$^{-1}$}
\usepackage{tikz}
\usetikzlibrary{positioning}
\usetikzlibrary{shapes,arrows,arrows,positioning,fit}

\author{Ali Al-Jaaidi}
\affiliation{Universit\'e Paris-Saclay, CNRS, Institut des Sciences Mol\'eculaires d'Orsay, 91405, Orsay, France.}
\author{David Lauvergnat}
\affiliation{Universit\'e Paris-Saclay, CNRS, Institut de Chimie Physique, 91405, Orsay, France.}
\email{David.Lauvergnat@universite-paris-saclay.fr}
\author{Daniel Pel\'aez}
\email{Daniel.Pelaez-Ruiz@universite-paris-saclay.fr}
\affiliation{Universit\'e Paris-Saclay, CNRS, Institut des Sciences Mol\'eculaires d'Orsay, 91405, Orsay, France.}

 \title{Variational computation of anharmonic ground and excited vibrational eigenstates using bound Quartic Force Fields:\\
 Application with MCTDH and ElVibRot}

\begin{document}

\maketitle

\begin{abstract}
In this work we introduce the use of Quartic Force fields (QFF) potential expansions in the context of variational calculations. Such potentials are commonly employed in molecular Vibrational Second-Order Perturbation Theory (VPT2) studies, for which equations explicitly dependent on the QFF parameters exist. However, QFF are unbound potentials for more or less large displacements from the reference point and, most of the time, this prevents their use in conjunction with variational wavepacket-based calculations. In this work, we propose a general correction to QFFs and introduce a fully automated numerical approach to avoid their unbound character. Our corrected potentials, bound QFF (bQFF), do not exhibit appreciable modification of the local topography around the region of interest for infrared spectroscopy. As a consequence of this, we can affirm that the vibrational eigenstructure (eigenvalues, eigenstates) remains essentially unaltered by our correction. To illustrate their numerical stability, we have interfaced our bQFF routines in combination with to two well-established quantum simulation software packages MCTDH and \textsc{ElVibRot} which feature variational approaches. More specifically, our bQFFs are separable and hence directly expressible as MCTDH operators. Furthermore, concerning the size of our bQFF expansion, we show that it is possible to tensor-decompose our bQFF in Canonical Polyadic form (CP-bQFF). We use the Monte Carlo Canonical Polyadic decomposition algorithm for this. CP-bQFF results are virtually identical to uncompressed bQFF, but the computational efficiency is largely improved. Our approach paves the way for the automated variational study of anharmonic eigenstates in molecular systems within the reach of QFF-based potentials, using either time-dependent or time-independent schemes. 
\end{abstract}

\section{Introduction}
Except for notable exceptions \cite{cam15:322}, the identification and assignment of chemical species from experimental vibrational spectra rely on the ability to accurately predict the molecular vibrational structure. This is more so in cases where recreating of the actual physical conditions is very hard, such as interstellar medium\cite{mac22:3198}, extreme environments,\cite{dar20:A82, xia25:11094}, as well as in the context of unstable or charged species, which require special experimental techniques which, in turn, may perturb the targeted signal.\cite{zen19:8, pel17:100, ger25:10339}

In this work by \emph{accurate prediction} (or simulation) we solely refer to methodologies (and associated software packages) which enable a variational quantum description of nuclei, such as the Multiconfiguration Time-Dependent Hartree Method (MCTDH)\cite{bec00:1} or \textsc{ElVibRot}.\cite{ElVibRot} In this respect, we also exclude the use of scaling factors (frequencies, intensities) except in the case of normalization or translation of the full spectrum. As the most prominent examples of these variational approaches, due to their relevance, one has the vibrational simulation of small water clusters\cite{sim25:184304, wan25:144308, larsson_computing_2019}, as well as their related charged counterparts\cite{ven05:104505, pel14:42, pel17:100, sch22:6170}

Fully general vibrational simulations are very hard tasks typically involving several research groups and the use of highly specialized methodologies.\cite{bow08:2145} Indeed, one needs to obtain a set of coordinates adapted to the energy regime and geometry of the system under consideration, the corresponding Kinetic Energy Operator (KEO)\cite{gat01:8275,lau02:8560} together with an accurate representation of the Potential Energy Surfaces (PES) and Dipole Operators (assuming a dipole mechanism). The latter quantities, in turn, involve accurate electronic structure calculations\cite{puz19:813} and cumbersome multidimensional fits, probably in the form of Deep Neural Networks\cite{qui18:262, nan24:8807, jia13:054112, fu23:nwad321, man22:10187, qiu26:xxx}. Arguably, the availability of a suitable PES is a major limiting factor. In this respect, it should be noted that some approaches might require a specific form (e.g., sum-of-products, SOP, as in some MCTDH implementations \cite {bec00:1, quantics:package}). In such cases, the PES should be already of separable form\cite{man06:194105, pan20:234110}, otherwise a further refit (i.e. tensor-decomposition) into a separable sum-of-products is necessary.\cite{jae96:7974, pel13:014108, sch17:064105, sch20:024108, qiu25:103873, qiu26:xxx} In contrast, some other methods are compatible with any type of potential expression (e.g. MCTDH using CDVR\cite{man26:214103}, \textsc{ElVibRot}\cite{ElVibRot},\textsc{TROVE}\cite{Yurchenko2007} or \textsc{GENIUSH}\cite{Matyus2009}.
Fortunately, such in-depth studies can be greatly simplified in the rather common case of semi-rigid molecules characterized by a single well. Indeed, for these, a local representation of the PES around a given minimum (assumed stable enough) suffices to determine its vibrational spectrum. Accurate local representations can be built using: (i) many-body type expansion\cite{sch22:174103}; (ii) adaptive methods guided by convergence of a molecular property (such as the Zero-Point Energy, ZPE)\cite{kli18:064113, ric26:5749}; or, an even simpler strategy, (iii) Taylor-type expansions of the energy up to a given order (fourth, sixth) around a reference point. The latter, usually referred to as Force Fields (FF) in spite of the obvious ambiguity of the term, have been cleverly exploited in the context of, probably, the most popular family of fully quantum vibrational simulations, the Vibrational Second-order Perturbation Theory (VPT2).\cite{bar04:014108, bar15:1301, fra21:1301} VPT2 is method of choice for the vast majority of users due to its relatively black box nature as epitomized by its successful application in the context astrochemistry, in spite of its intrinsic inability to deal with resonances such as Coriolis, Fermi and Darling-Dennison.\cite{can17:146} Note that it is possible to tackle such issues beyond a mere perturbative approach.\cite{men21:4332} For the sake of completeness, it should be noted that to overcome the aforementioned limitations, a hierarchy of approaches, in the same spirit of quantum chemistry VSCF, VCI, VCC, VMCSCF, VCASSCF,... exists.\cite{doi:10.1142/12305} Finally, it should be highlighted that the interest in molecular vibrational eigenstates goes beyond the mere computation of fundamentals but also determination of IR spectra as well as initial conditions for Molecular Dynamics\cite{AlisPaper}.

In this work, we will focus on semi-rigid molecules whose PES is given by a so-called Quartic Force Fields (QFFs).\cite{bar15:1301, wes23:2606} Note that sextic force fields in the case of small molecules have been successfully employed.\cite{cza04:23, for14:76, sib16:214107} Furthermore, we will assume that an appropriate choice of the level of level of electronic structure theory has been chosen\cite{puz19:813, for24:6528} and that the necessary numerical precautions have been taken into account.\cite{gei21:1018}  Quartic Force Fields (QFF) are Taylor expansions of a PES up to fourth order around a given reference position, $\mathbf{Q}_0$, usually an equilibrium geometry, and is given by: 
\begin{equation}\label{eq:qff}
    V^{\text{QFF}}(\mathbf{Q})=V_{\text{ref}}+
    \frac{1}{2!}\sum_{i}^f\Phi_{ii} Q_i^2+
    \frac{1}{3!}\sum_{i \ge j \ge k}^f\Phi_{ijk}Q_iQ_jQ_k + \frac{1}{4!}\sum_{i \ge j \ge k \ge l}^f\Phi_{ijkl}Q_iQ_jQ_kQ_l
\end{equation}

Here $f$ represents the number of internal degrees of freedom (3N-6 or 3N-5), $Q_\kappa$ is a (dimensionless, vide infra) displacement along normal mode $\kappa$, and $\Phi_{ii}=\frac{\partial^2 V}{\partial Q^2_i}$, $\Phi_{ijk}=\mathcal{P}_{ijk}\frac{\partial^3 V}{\partial Q_i \partial Q_j \partial Q_k}$, $\Phi_{ijkl}=\mathcal{P}_{ijkl}\frac{\partial^4 V}{\partial Q_i \partial Q_j \partial Q_k \partial Q_l}$ are partial derivatives of the full potential, $V(\mathbf{Q})$, along the indicated modes ($i, j, k,\ldots$). $V_{\text{ref}}$ is the energy at the reference position, $V_{\text{ref}}=V(\mathbf{Q}_0)$ and $\mathcal{P}_{\{\mu\}}$ is the factor accounting for the total number of derivatives related by permutation of the $\{\mu\}$ set of indices. Note in passing that it is common to restrict to the this expression to the so-called semi-QFF, a truncation of the quartic terms considering only at most two-mode couplings.\cite{ramakrishnan2015semi}\\

\begin{figure}[H]
    \centering
    \includegraphics[width=0.6\textwidth]{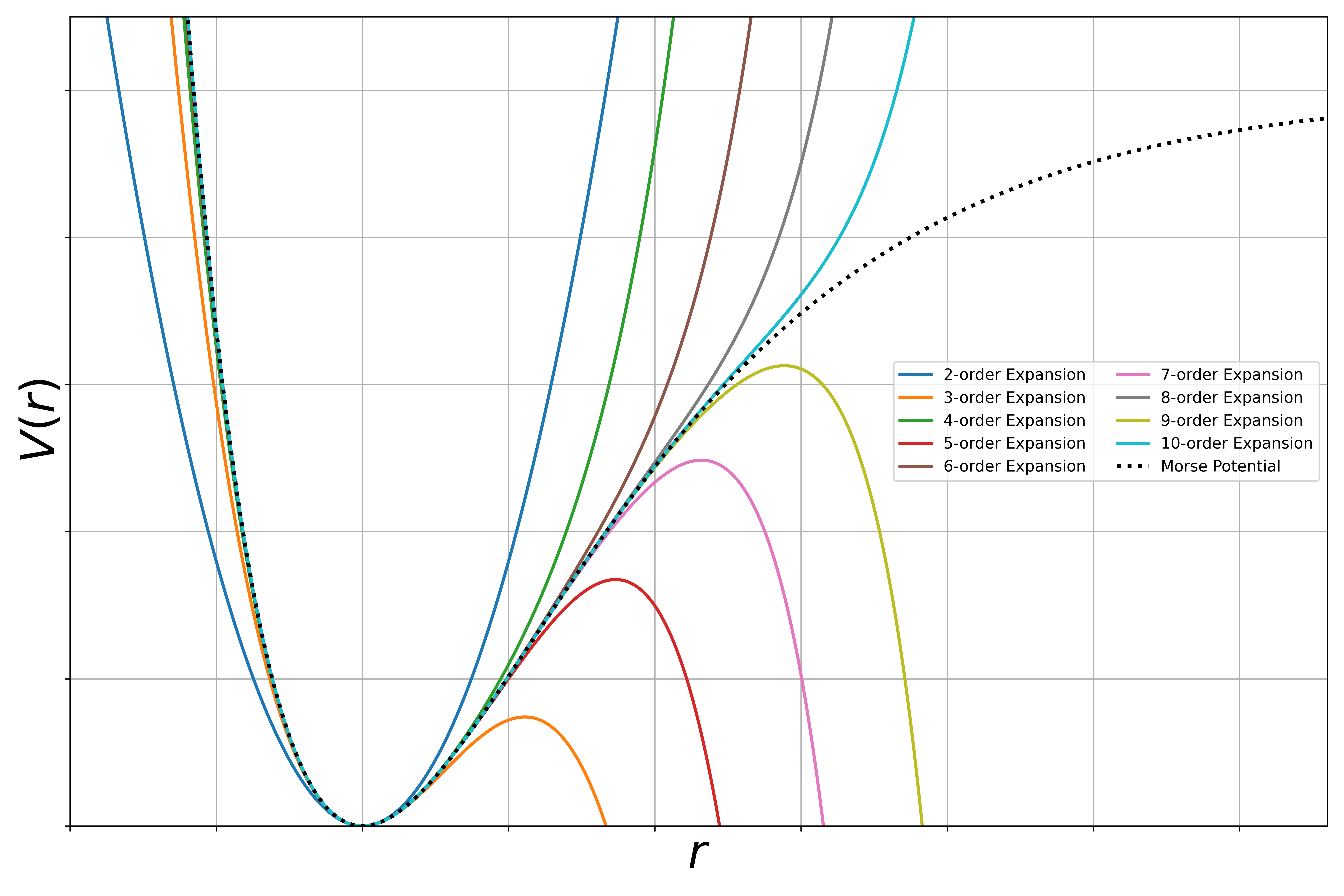}    
    \caption{Comparison of Taylor expansions of increasing order of a Morse potential (stretching mode) around its equilibrium geometry.}
    \label{fig:Taylor}
\end{figure}

As evident from their definition, QFFs are unbound potentials, and this prevents their universal application in combination with variational methods, which are at the core of our interests. Indeed, while harmonic frequencies are strictly positive for minima, the cubic terms and/or the quartic ones for $\Phi_{ijkl}<0$ and large displacements may become (largely) negative, i.e. unbound, hence exhibiting \emph{holes}. To illustrate this, in Figure ~\ref{fig:Taylor} we compare a Morse potential with its Taylor expansions up to 10$^{th}$ order.

\begin{figure}[H]
    \centering
    \includegraphics[width=0.6\textwidth]{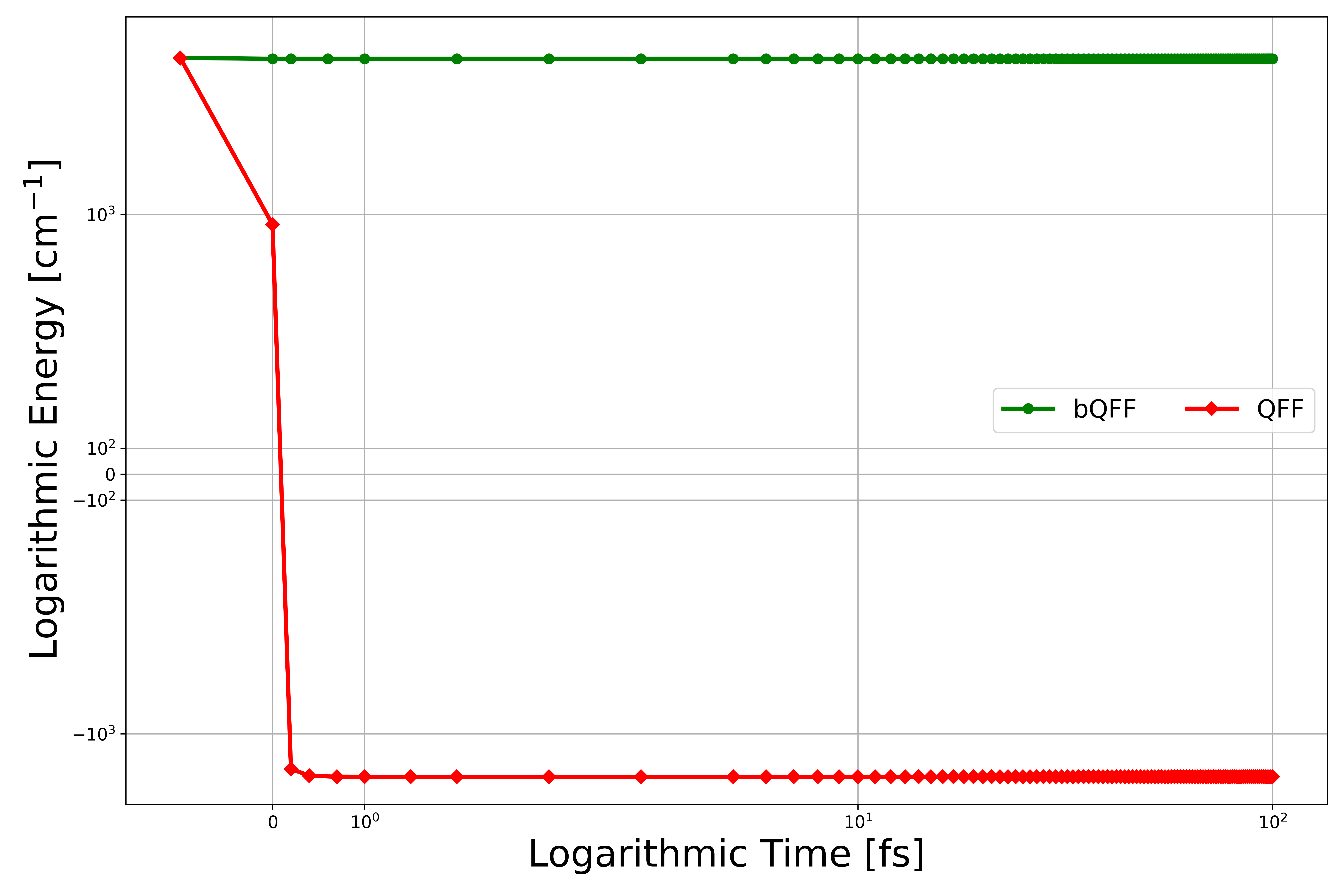}    
    \caption{Wavefunction relaxation\cite{kos86:223, mey06:179} on a bound and an unbound QFF water Potential.}
\label{fig:ShowingHoles}
\end{figure}

As a consequence, any variational quantum simulation, using either time-dependent or time-independent methods, is doomed to failure since regions of configuration space associated to such unphysically distorted unbound QFF will be eventually populated. This is epitomized in the context of vibrational spectroscopy\cite{pel13:014108, pel14:42, pel17:100}. To illustrate this, we present below a trivial example: the \emph{relaxation} of the ground state (GS) of the water molecule. 
Obviously, the expected result is a constant value, the Zero-Point Energy of water. In Figure ~\ref{fig:ShowingHoles}, we compare the behavior of the relaxation energy with time for two different QFF potentials, namely: (i) an unbound one (red, lower curve at long times) and (ii) a bound one (green, upper curve). Note that our graph uses a double logarithmic scale. In both simulations, the same initial wave function, built as a product of the 1D anharmonic ground states was employed. In the upper curve (green), the bound PES shows the expected behavior. In contrast, for the case of the unbound potential, a sudden drop in the energy towards negative (non-physical) values below the reference energy, $V_{\text{ref}}$) is observed, thus signaling the presence of a hole. Both potentials are reported in the Supplementary Information (SI).
 
For a QFF (or any high-order polynomial expression), detecting holes by, for instance, finding polynomial roots as a signature of a hole, is extremely cumbersome. To avoid such explorations, it is common to resort to simple solutions such as applying potential cuts in terms of energy or coordinate value.\cite{10.1063/1.2364892} Recently, Poirier and coworkers\cite{billfirst, billsecond} have introduced more systematic approaches to determine the actual locations of holes in full configuration space or in a subset of it (up to 30 DOFs).

In this work, we propose an automated system-independent procedure for the smooth (continuous, differentiable) correction of QFFs in semi-rigid systems, irrespective of their size. We denote the resulting PES as bound QFF (bQFF). To the best of our knowledge, no general correction method for  a PES presenting holes have been presenting before. In a nutshell, our algorithm relies on analyzing the couplings among the various modes at different orders and applying separable corrections that preserve the description of the original vibrational levels up to combination bands while rendering the PES bound at large displacements. This effectively removes holes thus enabling its interface with variational methods, in general, as well as a symbolic expression as in the case of an MCTDH operator. 

The remaining part of the article is structured as follows: in the Methodology Section, we describe all algorithmic and numerical details of our approach, more specifically, the generation of a bQFF, its efficient compression in Canonical Polyadic form, and all aspects concerning electronic structure calculations for our case studies. Section Results and Discussion presents two applications, a benchmark system which serves to illustrate straightforwardly our methodology and a full-dimensional study in 9D. Finally, in the Appendix we provide a detailed description of our quantum dynamical approaches and in the Supplementary Information we present extra data and graphs supporting our results.

\section{Methodology}
In this Section, we provide a description of the main methodological approaches employed in this work. First, we discuss the local approximation to molecular PES that we use, the so-called Quartic Force Fields. Among other aspects, we address their unbound character, which prevents their widespread use in combination with variational methods. Then we introduce a separable, parametric, correction to this issue, leading to bound Quartic Force Fields (bQFF). Finally, we show that such expansions can be efficiently compressed without appreciable impact on the associated vibrational structure. A succinct description the variational computational methods for the determination of vibrational eigenstates is provided in the Appendices. 

In what follows, unless otherwise stated, we will employ dimensionless normal coordinates. To obtain them, for a given reference geometry, one needs to diagonalize the mass-weighted Hessian ($\tilde{\mathcal{H}}$) of order 3N, N being the number of atoms. This yields a set of 3N eigenvalues corresponding to the squared angular frequencies ($\mathbf{w}$) and 3N associated eigenvectors ($\mathbf{D}$):
\begin{equation}
\mathbf{D}\tilde{\mathcal{H}}\mathbf{D}^T = \mathbf{w}
\end{equation}
Then 3N normal coordinates can be defined as the projection of the mass-weighted Cartesian displacements ($\Delta\tilde{x}$) on the eigenvectors as: 
\begin{equation}
    q_i=\sum_{\alpha}^{3N}D_{i\alpha }\Delta\tilde{x}_{\alpha} 
\end{equation}
Finally, transformation into dimensionless normal coordinates is done by frequency-scaling the normal coordinates as: 
\begin{equation}
\mathbf{Q}=\left\{\dots\sqrt{\frac{\omega_{i}}{\hbar}}q_{i}\dots\right\} \quad \forall i \in \{1, 2, \dots, f\}\label{eq:qandQ}
\end{equation}
where $f$=3N-6 for non-linear molecules. In this regard, it should be noted that the harmonic frequencies of non-vibrational modes are (ideally) zero. Consequently, dimensionless normal coordinates are only defined for the set of $f$ internal motions. Using dimensionless normal coordinates offers practical advantages for grid-based quantum dynamics calculations. In this representation, the primitive grid can be defined over the same numerical range for all modes, which simplifies its construction and the organization of the reduced density matrices. Once expressed in dimensionless form, their Hamiltonians share the same canonical structure, and differences in frequency are absorbed into the scaling. As a result, one can adopt common grid boundaries chosen large enough so that the wavefunction amplitudes are negligible at the edges for all modes, including moderately anharmonic ones.

\subsection{Bound Quartic Force Fields (bQFF)}

\begin{figure}[ht!]
    \centering
        \includegraphics[width=0.6\textwidth]{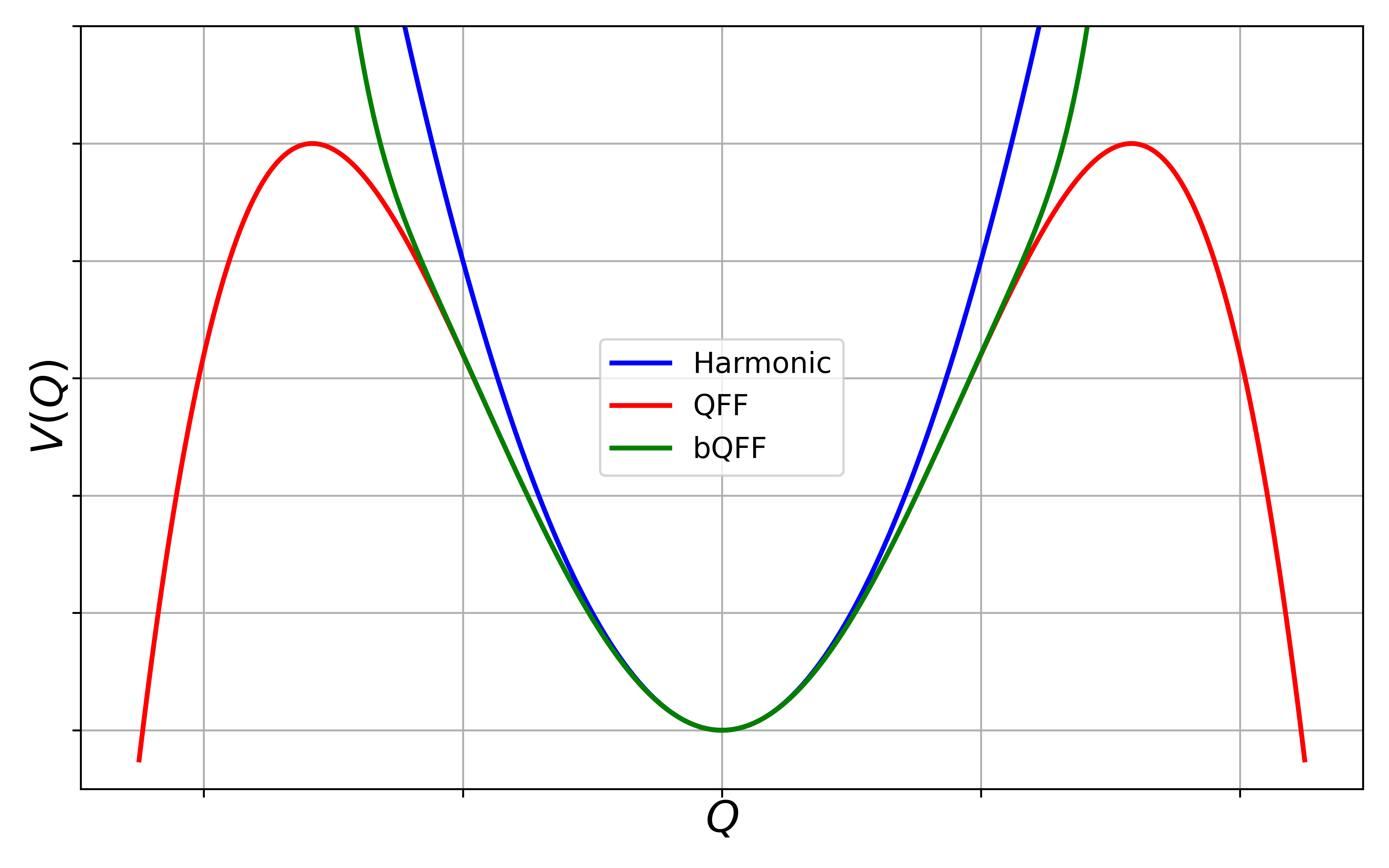}
    \caption{One-dimensional illustration of the bQFF to a simple quartic polynomial expression ($-(x^4-x^2)$).}\label{fig:AB}
\end{figure}

Ideally, a fully general method to plug the holes should satisfy three criteria. First, it should yield a reliable (faithful representation of the PES). Second, the corrected PES should be smooth. And third, as discussed above, in the case of MCTDH it should also yield a separable expression. In Figure \ref{fig:AB}, we compare a harmonic potential (blue, central curve) together with the associated QFF (red, unbound curve) and our bQFF (green, middle bound curve). The latter smoothly follows the QFF up to an \emph{inflection} region from which the QFF bounces back to negative values, while our bQFF smoothly and monotonically continues upwards in energy.

\begin{figure}[hb!]
    \centering
    \begin{subfigure}{0.45\textwidth}
        \centering
        \includegraphics[width=\linewidth]{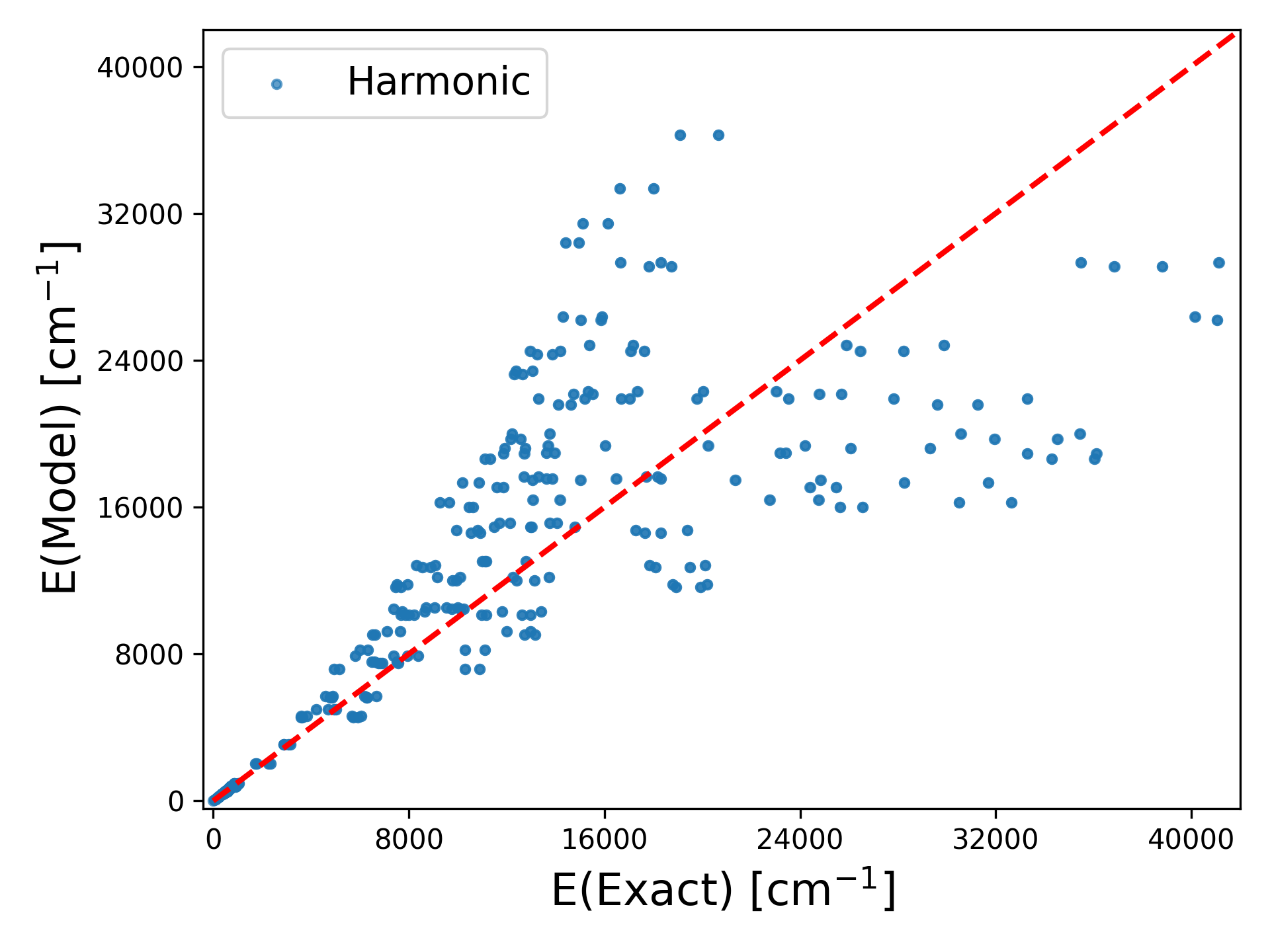}
        \caption*{(a)}
    \end{subfigure}
    \begin{subfigure}{0.45\textwidth}
        \centering
        \includegraphics[width=\linewidth]{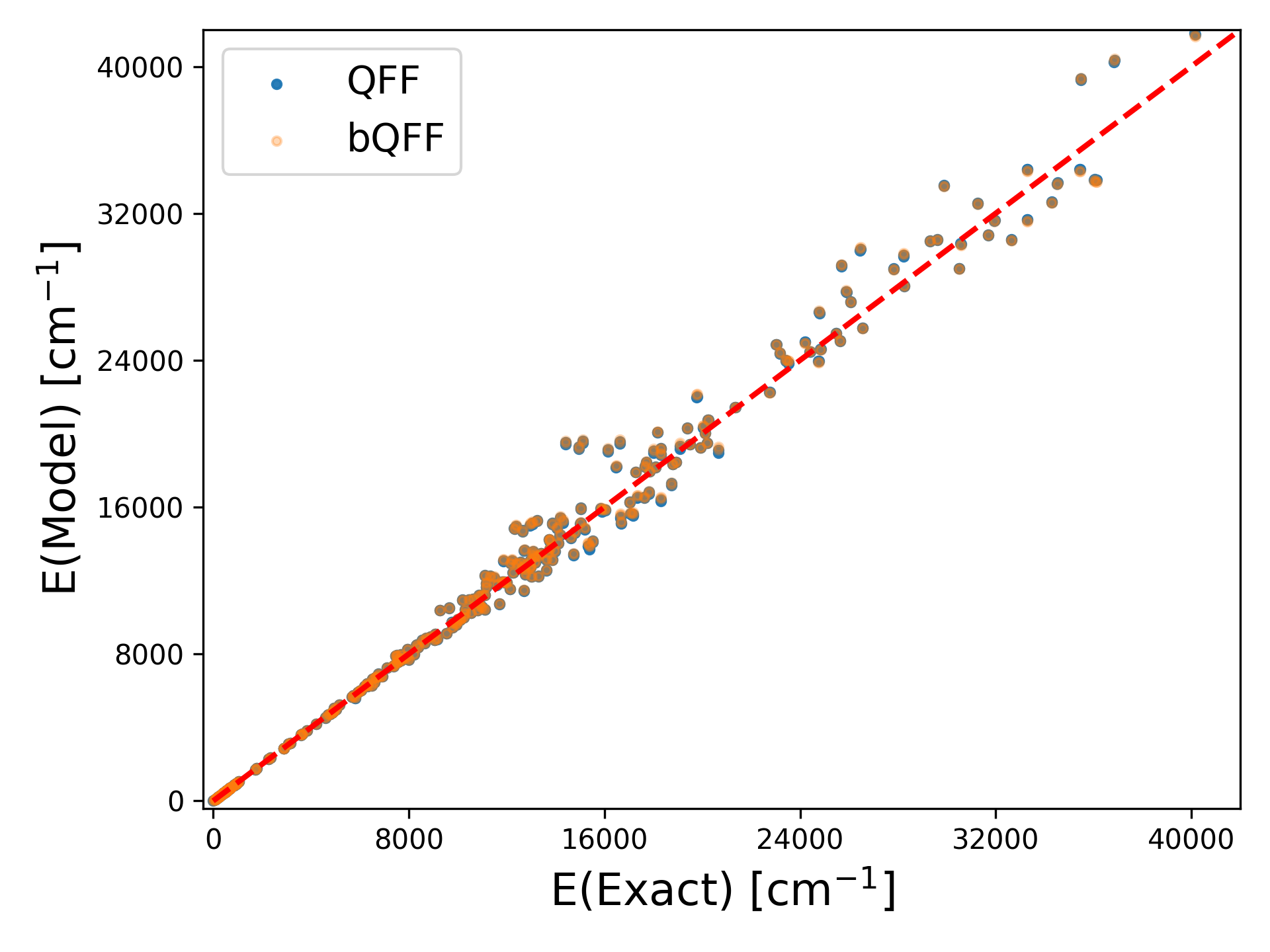}
        \caption*{(b)}
    \end{subfigure}

    \caption{Correlation plots of different water PES approximations: (a) harmonic and (b) QFF and bQFF.}
    \label{fig:AB2}
\end{figure}

This is also reflected in the error that can be visualized by means of correlation plots. In Figure \ref{fig:AB2} we present the correlation plots for different approximations to the ground state PES of water, namely harmonic, on panel (a) and QFF and bQFF on panel (b). In both panels, a red line ($y=x$) marks the exact PES value (from the reference electronic-structure calculation; see the Computational Details section). One can observe how the harmonic representation (panel (a)) rapidly deviates from the correct trend, whereas both QFF and bQFF (panel (b)) closely follow the same trend and yield much more controlled errors in a much  vaster energy range.

Whenever holes are present, common straightforward approaches include pointwise solutions such as filling the holes with spikes \cite {yu06:204306} or more dramatic ones such as cutting the potential according to certain criteria (energy, coordinates, etc.). In our appoach, bQFF, we effectively remove artifacts without identifying their exact locations and, at the same, providing a correction that is smooth and continuous over the entire configuration space.  

We will now discuss our approach to turning an unbound QFF potential into a bound potential (bQFF). Our solution simultaneously satisfies the three aforementioned criteria. As a QFF is a Taylor expansion of the PES up to fourth order, its applicability is for small to medium displacements. For large displacements, the QFF approximation to the energy is less and less accurate and not worth using in general. In the case of variational quantum calculations, such regions induce the incorrect behavior of the wave function (see discussion above).

Our corrected (bounded, b) QFF Hamiltonian is broken into two contributions of the form:
\begin{equation}\begin{aligned}
\hat{H}=\hat{H}_{\text{1D}}+ \hat{H}_{\text{Cross}}\end{aligned} \label{eq:splitted H}\end{equation}
finding holes in 1D cuts is straightforward. Our 1D corrected contribution (along a single mode) reads:
\begin{equation}\begin{aligned}
\hat{H}_{\text{1D}} = V_{\text{ref}} +\sum_{i=1}^f \Bigg [ \frac{\hbar\omega_i}{2} \bigg( - \frac{\partial^2}{\partial Q_i^2} + Q_i^2 \bigg) +\bigg(\Phi_{iii}Q_{i}^3+\Phi_{iiii}Q_{i}^4 \bigg) \cdot e^{-(\frac{Q_{i}}{Q_{i,\text{Hole}}})^{n1}} \Bigg]
 \end{aligned}\end{equation}
where $Q_{i,\text{Hole}}$ is the location of the maxima (automatically determined) or inflection point (if any) associated with mode $i$ after which the potential decreases, $e^{-(Q_i/Q_{i,\text{Hole}})^{n1}}$ is a window function with $n1$ and even number controlling its sharpness. Note that the 1D correction term will always be positive.

Regarding crossed terms, our correction is expressed as:
\begin{equation}\begin{aligned}
 \hat{H}_{\text{Cross}} = \frac{1}{3!}\sum^f_{i > j > k}\Phi_{ijk}Q_ie^{-\big(\frac{Q_i}{Q_\text{max}}\big)^{n_2}}Q_je^{-\big(\frac{Q_j}{Q_\text{max}}\big)^{n_2}}Q_ke^{-\big(\frac{Q_k}{Q_\text{max}}\big)^{n_2}}
\\+ \frac{1}{4!}\sum^f_{i > j > k > l}\Phi_{ijkl}Q_ie^{-\big(\frac{Q_i}{Q_\text{max}}\big)^{n_2}}Q_je^{-\big(\frac{Q_j}{Q_\text{max}}\big)^{n_2}}Q_ke^{-\big(\frac{Q_k}{Q_\text{max}}\big)^{n_2}}Q_le^{-\big(\frac{Q_l}{Q_\text{max}}\big)^{n_2}} 
\end{aligned}\end{equation}
in analogy to the previous 1D window functions, we multiply each coordinate $Q_i$ in each crossed term by a window function $e^{-(Q_i/Q_\text{max})^n2}$. Here, $Q_\text{max}$ delimits the region to be corrected and corresponds to the maximum displacement used to scale crossed terms, with $n_2$, also an even number, controlling its sharpness. 

Choosing the $Q_\text{max}$ values is rather intuitive and system-dependent. As a first guess, one can simply consider vanishing populations of the corresponding harmonic oscillator fundamental, for instance, wavefunction for increasing values of the displacement $Q$. In other words, the correction should be applied in those regions where the wavefunction population for any state of interest is vanishingly small. Note that if no holes are present, then $Q_\text{max}=\infty$. On the other hand, $n_1$ (wherever needed) and $n_2$, should be chosen by the user. Our experience seems to indicate that the correction is rather system insensitive. In this work, all calculations employ the same values of $n1$ (when needed) and $n2$ (see the operator files in the Supporting Information). Note that the stability of these parameters has been thoroughly tested (see Supporting Information).

Concerning the behavior of bQFF on the fundamentals, it should be highlighted that the 1D corrections preserve the exact QFF potential up to $Q_{i,\text{Hole}}$ and, hence, do not impact their representation. However, the crossed terms may have a larger effect whenever $Q_\text{max} < Q_{i,\text{Hole}}$. In our specific study-cases, we have not observed appreciable impact of the corrections on the fundamentals or on the combination bands.

In sum, bQFF provides a stable bound and representation for a QFF, with a good trade-off between accuracy and computational cost. Our current bQFF code provides an analytical and separable expression of a PES as needed in a an MCTDH operator file. It is also a callable subroutine that can be interfaced with MCCPD to obtain CP-bQFF (see next section) or used directly with a quantum dynamical software package such as \textsc{ElVibRot}. 

\subsection{Canonical Polyadic decomposition of Bound Quartic Force Fields (CP-bQFF)}

As discussed above, our bQFF are fully general in the sense that their analytical and separable nature renders them compatible with virtually \emph{any} software package. This is particularly true for the Heidelberg MCTDH and Quantics packages, since the bQFF can be expressed directly in symbolic form as an $\texttt{operator}$ file (see Appendix). However, the number of QFF (or bQFF) terms ($N_{\text{QFF}}$) increases significantly with the number of active degrees of freedom ($f$). Specifically, there could be $\frac{f(f+1)(f+2)}{6}$ and $\frac{f(f+1)(f+2)(f+3)}{24}$ cubic and quartic terms, respectively. In Table \ref{tab:QFF Size}, we provide some illustrative examples of this growth. Note that for some systems or applications, the actual number of terms may be reduced by application of a numerical threshold.

\begin{table}[h]
\centering
\begin{tabular}{c c c }
\hline
Molecule  & Dimensionality & $N_{\text{QFF}}$ \\
\hline
Water & 3D & 28 \\
CHFClBr & 9D & 669 (461)\\
Pyrazine & 24D & 20174 (1417) \\
Naphthalene & 48D & 269548 (10080) \\ 
Pentacene & 102D & 4962436 (350473) \\
\hline
\end{tabular}
\caption{Increase in the number of QFF parameters (harmonic frequencies, cubic and quartic terms) with the number of dimensions. The third column presents the number of unique terms and, in parenthesis, the numerically relevant ones.}
\label{tab:QFF Size}
\end{table}

Owing to this growth, storing and operating such expansions, although doable, might become computationally inefficient. To alleviate this, we have re-expressed our bQFFs in the much more compact Canonical Polyadic (CP) form. Our specific choice of CP algorithm is the Monte Carlo Canonical Polyadic Decomposition (MCCPD), implemented in MCTDH.\cite{sch20:024108} The  CP-bQFF expression for the geometry associated to grid point ($i_1\ldots i_f$) reads:
\begin{equation}
    V^{\text{bQFF}}_{i_1\ldots i_f}\approx V^{\text{CP-bQFF}}_{i_1\ldots i_f} = \sum_{r=1}^{R}\lambda_r \prod_{\kappa=1}^f v^{(\kappa)}_{i_\kappa, r}
\end{equation}
where $R$ is the rank of the CP decomposed tensor. As it will be shown below, this transformation leaves the potential and, concomitantly, the associated vibrational spectrum essentially unaltered.

\subsection*{Computational details}\label{sec:comp}
All of our equilibrium geometries and associated QFF expansions have been obtained using the Gaussian 16 Software package\cite{g16}. Reference geometries have been determined using the $\texttt{opt=verytight}$ option.
The complete headers of the input files are provided in The Supporting Information. Concerning the level of electronic structure theory, the QFF for water has been obtained at the CAM-B3LYP+D3BJ/Def2TZVP level of theory. For CHFClBr/CDFClBr molecules, we have employed the MP2/aug-cc-pVTZ level of theory. The determination of vibrational eigenstates was carried out using bQFFs interfaced to two different software : (1) Gaussian 16 GVPT2 implementation (hereafter referred to as VPT2) and (2) two variational methods from: (i) the Heidelberg MCTDH package and (ii) the \textsc{ElVibRot} package. The computational strategies for this are described in Appendices \ref{app:mctdh} and \ref{app:elvibrot}, respectively. 

\section{Results and discussion}
We will now illustrate the performance of bQFF with two molecular systems, namely: (i) the (3D) water molecule, which will serve as a benchmark; and (ii) the 9D anharmonic CHFClBr system together with its deuterated counterpart. It should be clear that our goal is not to achieve the best possible vibrational eigenvalues but to show that our bQFF potentials can be used in conjunction with variational methods (including wave packet propagation, \emph{vide infra}) yielding accurate and numerically converged ground state (GS) energies and vibrational eigenstates (fundamentals and combination bands). The quality of these values will be solely limited by the choice of the underlying level of electronic structure theory (method and electronic basis set) through the force constants (potential derivatives) and the chosen expansion order. An extensive comparison of the variational results using our bQFF and the uncorrected QFF is presented in the Supporting Information (SI). Regarding the behavior of variational methods with QFFs, it should be noted that even in the case of PES with holes, convergence might be achieved with a bunch of strategies carefully selecting: (i) primitive grid density and boundaries\cite{pel13:014108}, (ii) an appropriate number of basis functions or single particle functions, and (iii) a suitable initial wavefunction. Note that in the context of the Density Matrix Renormalization Group approach the so-called state-shifting technique has been shown to efficiently handle such issues\cite{lar25:3991}. In general, one can affirm that the straight use of QFF potentials requires the aforementioned precautions together with a careful assessment of the results. Obviously, such strategy lacks generality, and each new system would require a full independent study with no guarantee of success. In contrast to this, achieving a bQFFs is fully automated and solely depends on the Taylor expansion ansatz and, as such, is system independent. 

\subsection{Proof of concept and benchmark: 3D water molecule}
A QFF expansion for the water molecule was obtained as described in Section \ref{sec:comp}. The resulting expansion comprises 28 anharmonic terms. The corresponding numerical parameters are provided in the Supplementary Information. VPT2 and bQFF GS together with selected vibrational energy levels are reported in Table~\ref{tab:water modes}. The difference between the two sets has two origins: the different methodologies (perturbative and variational) and the inclusion of Watson-Coriolis coupling in VPT2, which contributes approximately $-11~\text{cm}^{-1}$ to the ZPE. This contribution is intentionally excluded from our method, 
as it is designed for larger molecular systems where such effects are typically negligible. This assumption will be shown to be valid in our subsequent CHFClBr/CDFClBr study (Section \ref{sec:fullD}). Nevertheless, to show that such differences cannot be attributed to our correction in SI-Table 3-5, we present results using the uncorrected QFF, which are virtually identical to our bQFF. Small deviations may nevertheless appear for certain vibrational states, a behavior due to the fact that bQFF is a correction that depends on the actual value of the coordinate. Hence, shifts relative to the uncorrected QFF are expected to arise primarily for vibrational states that imply large displacements, such as higher-frequency overtones. In contrast, ground state, fundamental bands, and even combination bands are essentially unaffected.

Table \ref{tab:water modes} displays a comparison between harmonic, VPT2, bQFF and a canonic polyadic (CP) decomposition of our bQFF hence denoted as CP-bQFF. The latter has been obtained with the so-called Monte Carlo Canonical Polyadic Decomposition (MCCPD) approach using a uniform sampling over the primitive grid.\cite{sch20:024108}

\begin{table}[h]
\centering
\begin{tabular}{c c c c c }
\hline
State  & Harmonic & VPT2  & bQFF  & CP-bQFF  \\
\hline
ZPE & 4683.692 & 4614.235 & 4631.549 & 4631.548\\
$\nu_1$ & 1611.109 & 1561.220 & 1538.709 & 1538.709 \\
$\nu_2$ & 3826.061 & 3664.675 & 3700.380 & 3700.380 \\
$\nu_3$ & 3930.214 & 3755.805 & 3784.157 & 3784.157 \\
$\nu_1$ + $\nu_2$ & 5437.169 & 5215.461 & 5199.443 & 5199.443 \\
$\nu_1$ + $\nu_3$ & 5541.323 & 5301.738 & 5255.381 & 5255.381 \\
\hline
\end{tabular}
\caption{Comparison of the eigenenergies associated to selected low-lying vibrational eigenstates of water computed at different levels of theory.}
\label{tab:water modes}
\end{table}

As it can be observed, even in this unfavorable case (see discussion on Watson-Coriolis coupling above), both bQFF and CP-bQFF fundamental frequencies are in good and, more importantly, consistent agreement with the VPT2 values. More relevantly, for selected combination bands, our variational results are consistently lower than the perturbative VPT2 ones.

\subsection{Full-dimensional (9D) vibrational study of \ce{CHFClBr} and \ce{CDFClBr}}\label{sec:fullD}
\begin{figure}[H]
    \centering
\includegraphics[width=0.6\textwidth]{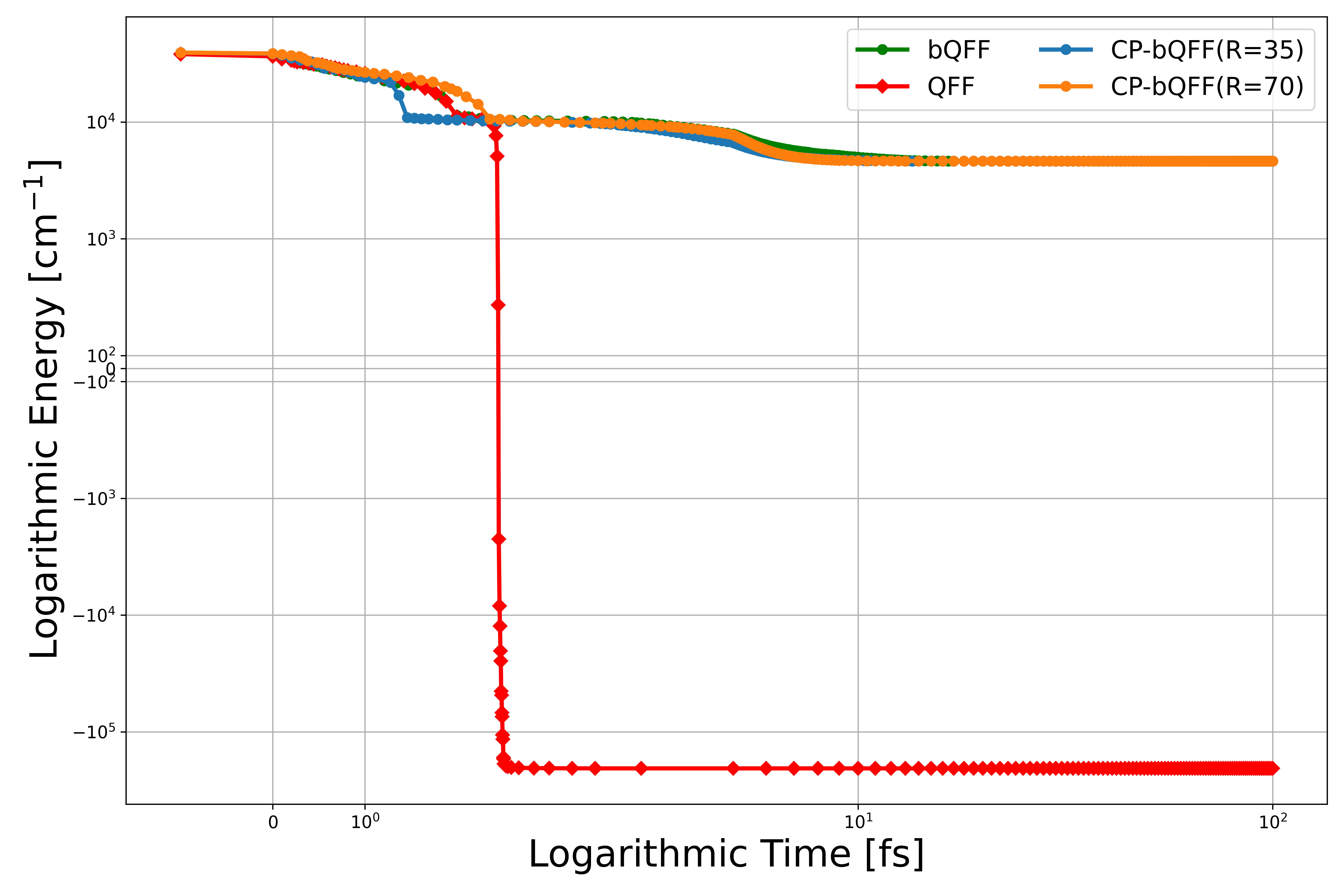}    \caption{Comparison of the stability of QFF and bQFF in reaching the hole regions using an excited initial wavepacket.}
    \label{fig:Holes CHFClBr}
\end{figure}

As mentioned, the Watson-Coriolis term can be neglected for large system or systems with relatively heavier atoms such as CXFClBr (X=H or D). For these systems, the Watson-Coriolis contribution of the ZPE is 3 orders of magnitude smaller than in water. This system was previously studied by Rauhut \emph{et al.} using VPT2 and VCI\cite{rauhut2006vibrational} as well as VCI and VSCF\cite{rauhut2021parity}. Our results are displayed in Table~\ref{tab:CDFClBr modes}. As expected, both the ZPE and all the fundamental bands are reproduced by bQFF with differences of around $1~\text{cm}^{-1}$ when compared to VPT2, and practically identical results to QFF (see SI-Table 5). Note that the anharmonic character of the deuterated species is not very significant thus explaining the very good agreement between the two methods. Once shown that our correction does not perturb the VPT2 results significantly, we can investigate its behavior in a system with an increase in the anharmonic character (setting X=H).\\
\begin{table}[h]
\centering
\begin{tabular}{ c c c c c }
\hline
State  & Harmonic & VPT2  & bQFF & Expt.$^{\text{*}}$  \\
\hline
ZPE & 3890.319 & 3860.463 & 3860.264 & - \\
$\nu_1$ & 229.370 & 227.794 & 227.709 & 222.7\\
$\nu_2$ & 320.091 & 317.054 & 316.905 & 312.7\\
$\nu_3$ & 430.085 & 425.831 & 425.694 &  423.2\\
$\nu_4$ & 644.208 & 638.824 & 638.445 & 619.9\\
$\nu_5$ & 768.436 & 756.198& 756.464 & 748.4\\
$\nu_6$ & 948.931 & 929.038 & 928.237 & 917.9\\
$\nu_7$ & 989.699 & 973.428& 972.827 & 974.3\\
$\nu_8$ & 1099.348 & 1072.615 & 1072.948 & 1082.7\\
$\nu_9$ & 2350.470 & 2286.792 & 2288$\pm$1$^a$ & 2265.3 \\
\hline
\end{tabular}\\
$^{\text{*}}$:Data taken from Ref\cite{beil2000vibrational}.\\
$^a$: see discussion on $\nu_9$ in the CHFClBr section.
\caption{Vibrational Modes of CDFClBr}
\label{tab:CDFClBr modes}
\end{table}
\FloatBarrier
The computed vibrational energy levels for CHFClBr are reported in Table~\ref{tab:CHFClBr modes}. A clear shift in energies is observed compared to CDFClBr as the heavier deuterium increases the reduced mass of the vibrating bond, leading to a decrease in vibrational frequencies. As a result, all the fundamental C–D stretching and bending modes appear at lower wavenumbers compared to the corresponding C–H modes. We also report the vibrational levels using the uncorrected QFF potential in SI-Table 4.\\
As mentioned at the beginning of this section, it might be confusing how one can get the low-lying vibrational levels using the uncorrected QFF, since it has holes. For the ground states our system, the answer is simply the initial wavepacket. Of course, the initial wavepacket should be close to the desired eigenstate. This can be readily done by constructing an initial wavepacket as a product of the 1D anharmonic (QFF, in our case) eigenstates. In our case, we manage to converge the vibrational GS and some fundamentals even in the presence of holes (unlike in the case of water, see Supporting Information). To show how this is a lucky case of QFF, we simply constructed the initial wavepacket as product of excited 1D eigenstates while keeping everything else exactly the same. We show the results in Figure \ref{fig:Holes CHFClBr}. As expected, we see that bQFF comes back exactly to the GS in Table \ref{tab:CHFClBr modes} while the QFF gives nonphysical results. This shows how reliable bQFF can be in plugging the holes, while persevering QFF topography.

 By comparing VPT2 and bQFF, one can observe that the GS and the first six fundamental bands are very close in agreement, whereas noticeable deviations are observed for the $\nu_7$ and $\nu_8$ modes. Interestingly, such deviations are not due to the bQFF correction but to the perturbative nature of the VPT2 method. As in the case of water, we show in SI-Table 4 that the ZPE and the first 8 fundamentals are essentially unchanged between QFF and bQFF. Instead, while our approach can be variationally converged, this is not guaranteed for a second-order perturbation theory like VPT2, this is particularly so  since $\nu_7$ it is known to exhibit a strong resonance with $\nu_5 + \nu_3.$\cite{beil1996fermi} 
 
\begin{table}[h]
\centering
\begin{tabular}{ c c c c c c}
\hline
State  & Harmonic & VPT2  & bQFF/MCTDH &bQFF/\textsc{ElVibRot} & Expt.$^{\text{*}}$  \\
\hline
ZPE & 4676.154 & 4628.330 & 4627.474 & 4627.474 & -\\
$\nu_1$ & 230.355 & 228.476 & 228.369 & 228.371&223.6\\
$\nu_2$ & 320.810 & 317.555 & 317.349 &317.352 &313.0\\
$\nu_3$ & 432.326 & 427.916 & 427.774 & 427.776&425.2\\
$\nu_4$ & 685.931 & 678.278 & 677.951 &677.952 &663.6\\
$\nu_5$ & 809.526 & 794.246 & 794.717 &794.718 &787.0\\
$\nu_6$ & 1095.124 & 1066.989 & 1068.422 &1068.43 &1077.2\\
$\nu_7$ & 1248.613 & 1226.018 & 1212.455 &1212.43 &1202.8\\
$\nu_8$ & 1339.156 & 1309.824 & 1304.720 &1304.72 &1306.2\\
$\nu_9$ & 3190.468 & 3059.131 & 3061$\pm1 $ & 3061$^a$ & 3025.5\\
\hline
\end{tabular}\\
$^{\text{*}}$:Data taken from Ref\cite{beil1996fermi}.\\
$^a$: Level computes with a smaller Smolyak basis set $L=6$ and its accuracy is reduced (difference between levels computed at $L=5$ and $L=6$).
\caption{Vibrational Modes (in \icm) of CHFClBr}
\label{tab:CHFClBr modes}
\end{table}

However, the $\nu_9$ region requires a more detailed discussion. Indeed, due to the high density of states around 3100 \icm, the convergence of the $\nu_9$ level is difficult. With our block Davidson procedure used in \textsc{ElVibRot}, more than 1000 states must be converged before converging the $\nu_9$ level for a Smolyak basis parameter, $L$, equal to $6$. This prevents converging the $\nu_9$ level with a large Smolyak basis set ($L=7$). With the improved relaxation technique implemented in Quantics/MCTDH, the difficulty of converging the $\nu_9$ level is similar with lock relaxation. To tackle this in both systems, we have propagated with MCTDH an initial wavepacket excited around $\nu_9$ for long times up to 6.5 ps until a single peak (width $\pm 1 cm^{-1}$). The value reported corresponds to the center of that peak.\\

\begin{figure}[H]
    \centering
    \includegraphics[width=0.8\textwidth]{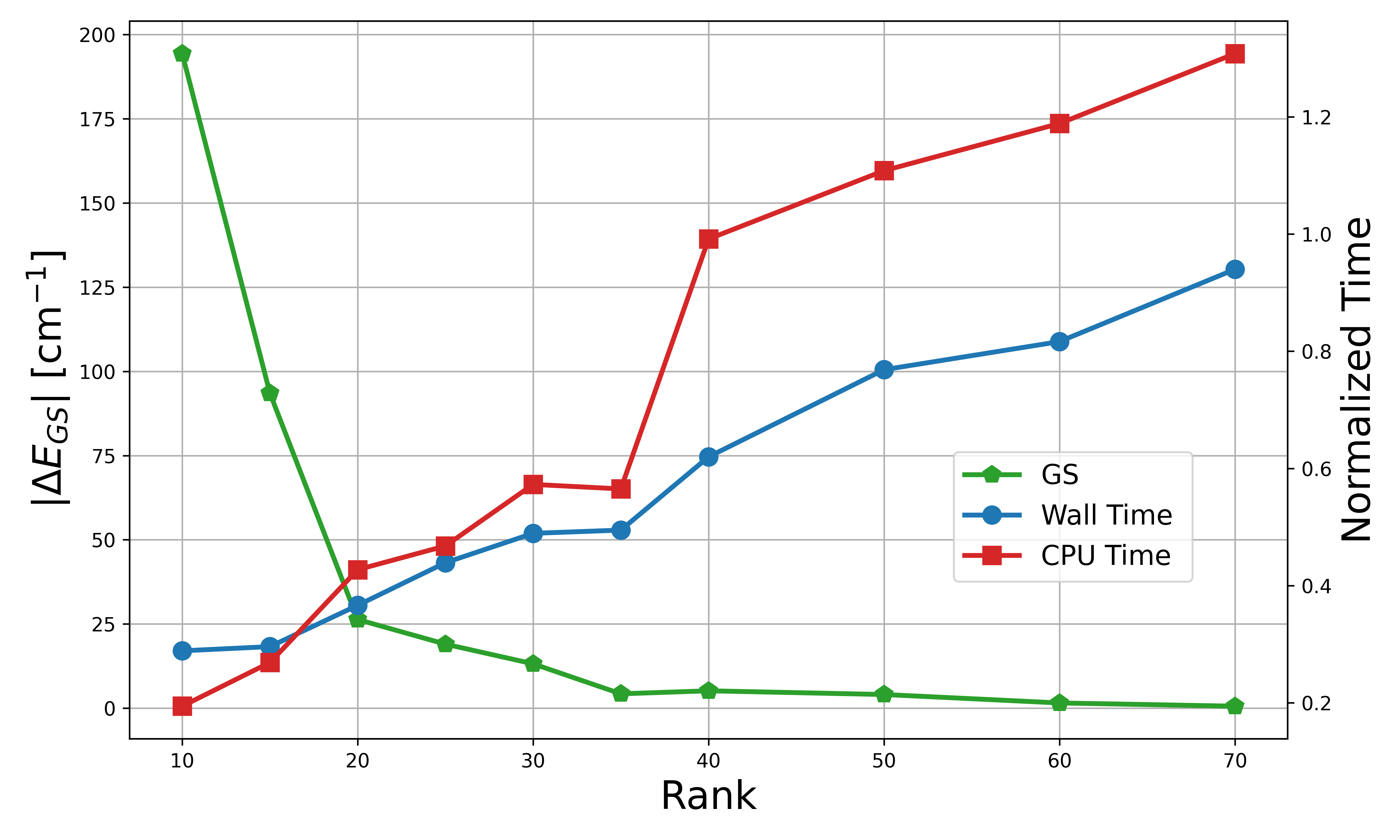}    
    \caption{GS Error and CPU Times vs. Rank}
    \label{fig:GSErrorTime}
\end{figure}

To finalize this Section, we note that the bQFF expansions can be expressed more compactly using MCCPD.\cite{sch20:024108}  Figure~\ref{fig:GSErrorTime} reports the (GS) error relative to bQFF as a function of the CPD rank, together with the corresponding wall- and CPU-times. It is observed that a CPD rank of 70 yields a negligible GS error ($<$ 0.5 \icm) while requiring roughly the same wall-time as the full bQFF calculation. The RMSE for such a rank is 8.4 \icm. The strength of CP-bQFF lies in the fact that, although increasing the CPD rank leads to an approximately linear increase in wall-time, the accuracy of the tensor fit—and consequently the reduction in GS error—improves much more rapidly (exponential-like). For instance, at a CPD rank of 35, the GS error is approximately $4~\text{cm}^{-1}$, while the wall time is reduced by about 50\%. This makes CP-bQFF a very good candidate for studying the vibrational structure of high-dimensional anharmonic systems. 
The vibrational frequencies of CHFClBr computed using CP-bQFF at ranks 35 and 70 are reported in Table~\ref{tab:CHFClBr modes with CP}. For simplicity reasons, we only show the first 8 fundamentals which are the ones that can be converged using the lock relaxation scheme in MCTDH. At the lower rank, aside from a $4~\text{cm}^{-1}$ shift in the GS energy, CP-bQFF yields reliable vibrational frequencies while requiring, on average, approximately 50\% less computational time than the full bQFF calculation. In contrast, the higher-rank CP-bQFF representation essentially reproduces the full bQFF results. The remaining minor discrepancies are attributable to sampling effects, as only a small fraction of the $19^9$ tensor points were evaluated in constructing the CPD tensor. 
\begin{table}[h]
\centering
\begin{tabular}{ c c c c }
\hline
State  & CP-bQFF & CP-bQFF  & bQFF   \\
  & $R=35$ & $R=70$  &    \\
\hline
ZPE & 4631.742 & 4628.070 & 4627.474 \\
$\nu_1$ & 227.453 & 227.830 & 228.369 \\
$\nu_2$ & 317.501 & 317.469 & 317.349 \\
$\nu_3$ & 426.961 & 427.710 & 427.774 \\
$\nu_4$ & 676.801& 677.592 & 677.951 \\
$\nu_5$ & 793.991 & 794.905 & 794.717 \\
$\nu_6$ & 1067.314 & 1068.586 & 1068.422 \\
$\nu_7$ & 1213.266 & 1212.608 & 1212.455 \\
$\nu_8$ & 1305.262 & 1304.964 & 1304.720 \\
\hline
\end{tabular}
\caption{Vibrational Modes (in \icm) of CHFClBr for Different CP Ranks}
\label{tab:CHFClBr modes with CP}
\end{table}
Similarly, for CDFClBr, CP-bQFF reproduced those results with an average of 50\% reduction in computational time, demonstrating a more compact structure. The same  $4~\text{cm}^{-1}$ ZPE shift is observed as shown in Table \ref{tab:CDFClBr modes with CP}. To finalize our discussion on CP-bQFF, it should be highlighted that the actual CP decomposition does not seem to significantly modify the structure of the underlying bQFF. More specifically, we do not observe the creation of artificial holes as shown in Figure \ref{fig:Holes CHFClBr} and, as discussed, the eigenvalues are almost identical (see Tables \ref{tab:CDFClBr modes with CP} and \ref{tab:CHFClBr modes with CP}).
   
\begin{table}[h]
\centering
\begin{tabular}{ c c c c }
\hline
State  & CP-bQFF & CP-bQFF  & bQFF   \\
  & $R=35$ & $R=70$  &    \\
\hline
ZPE & 3854.757 & 3858.996 & 3860.264 \\
$\nu_1$ & 227.659 & 227.719 & 227.709\\
$\nu_2$ & 317.598 &317.112 & 316.905\\
$\nu_3$ & 425.236 &426.143 & 425.694\\
$\nu_4$ & 640.709 & 638.624& 638.445\\
$\nu_5$ & 755.782 & 756.808& 756.464\\
$\nu_6$ & 931.040 &928.740 & 928.237\\
$\nu_7$ & 973.352& 973.193& 972.827\\
$\nu_8$ & 1072.233& 1073.224 & 1072.948\\
\hline
\end{tabular}
\caption{Vibrational Modes (in \icm) of CDFClBr for Different CP Ranks}
\label{tab:CDFClBr modes with CP}
\end{table}

\section{Data availability}
The data that support the findings of this study are available from the corresponding author upon reasonable request. Codes and examples will be available in the Heidelberg MCTDH (\url{https://www.pci.uni-heidelberg.de/tc/usr/mctdh/doc/index.html}) and Quantics (\url{https://www.chem.ucl.ac.uk/quantics/doc/}) software packages.

Furthermore, the \textsc{ElVibRot} code\cite{ElVibRot} used to perform the time-independent calculations is available on github and Zenodo\cite{ElVibRot_CHFClBr}. 
The CHFClBr bQFF and QFF potentials are implemented in the Quantum Model Lib\cite{QML} package available on github and Zenodo\cite{QML_CHFClBr}.\\

\section{Acknowledgements}
The work was supported by the Agence Nationale de la Recherche (ANR) SINGLETFISSION Grant No: ANR-23-CE29-0007 (\url{https://anr.fr/Projet-ANR-23-CE29-0007}). AAJ gratefully acknowledges his Ph.D. Grant from the ANR project ANR-23-CE29-0007. The authors gratefully acknowledge the financial support of the Institut des Sciences Moléculaires d'Orsay (ISMO), Institut de Chimie Physique (ICP) of the Université Paris Saclay and CNRS. AAJ and DP would like to thank kindly J.-Y. Bazzara, A. G. Borisov, L. Coudert and M. Leoni for their kind computational support with the different clusters at Paris-Saclay University. 
DL thanks the CNRS, France, for local computing facilities at ICP. D.P gratefully acknowledges fruitful discussions with B. Pouilly and M. Monnerville (Université de Lille).

\section{Conclusions}
In this work, we introduce an automated method for the generation bound quartic Taylor-series expansions, denoted as bound Quartic Force Fields (bQFF). Our results conclusively demonstrate that the proposed ansatz can efficiently correct the holes in the underlying QFF while preserving its overall curvature. This makes the resulting bQFF potential suitable for variational methods such as MCTDH and \textsc{ElVibRot}, enabling the computation of vibrational properties beyond a perturbative picture. Furthermore, the resulting potential is naturally expressed in seperable form as a sum-of-products (SOP), which is advantageous for MCTDH calculations. We demonstrate that it can be further compressed using tensor-decomposition techniques such as Monte Carlo Canonical Polyadic Decomposition. The resulting expression yields approximately a twofold speed-up in our variational calculations. Our method has been benchmarked against water (3D), and tested on CHFClBr and CDFClBr 9D systems. For moderately anharmonic systems, a clear agreement with VPT2 results is observed. For more strongly anharmonic ones, some deviations from the perturbative results may arise.

\clearpage
\section{Appendices}

\subsection{Determination of vibrational eigenstates with MCTDH}\label{app:mctdh}

As it is well-known,\cite{kos86:223} the lowest eigenstate of a system can be obtained by direct propagation of an initial guess wave packet in negative-imaginary time. Excited states can be also obtained through relaxation by imposing orthogonality to lower-lying eigenstates. This procedure might become cumbersome if na\"ively employed. Hence, in MCTDH a modification thereof was introduced: the so-called Improved Relaxation.\cite{mey06:179} An MCTDH wave function is propagated as in relaxation, but using equations of motion which are derived using a time-independent
variational principle.\cite{mey03:251,mey06:179} As main difference with relaxation, the so-called $\mathcal{A}$-vector
is obtained by diagonalization of the Hamiltonian matrix in the basis of the (current) configurations. The procedure is repeated until convergence. A Block Improved Relaxation algorithm exists, in which a common set of single-particle functions are relaxed using state-averaged mean-fields and a block of eigenstates is determined by block-Davidson
diagonalization.\cite{dor08:224109} In this framework, the computation of specific eigenstates can be achieved by generating guess wavefunctions whose nodal structure is similar to the targeted eingenstate. For the case of fundamentals, such a guess function can be readily achieved by operating the position operator for the relevant mode $\kappa$, $\hat{q}_\kappa$, on the ground-state wavefunction:\cite{ven09:034308, pel14:42}  
\begin{equation}\label{ass1}
|\Phi_{test} \rangle = N \hat{q}_\kappa |\Psi_{GS} \rangle \ , 
\end{equation}
where $N$ is a normalization constant. In other words, our guess function differs from the GS in \emph{one quantum of excitation} on the $\hat{q}_\kappa$ mode. A good estimate of the character of the final eigenstate can be obtained by overlap with a series of test function(s).\cite{pel14:42}

\subsection{Determination of vibrational eigenstates with \textsc{ElVibRot}}\label{app:elvibrot}

On way to overcome the curse of dimensionality is the use the Smolyak scheme. Indeed, in 1963, Smolyak proposed a procedure to transform a single and large direct-product, $\mathbf{S}^{DP}_{\bm{\ell}}$ (Eq. \ref{Eq_DP}), to a selected sum (or union) of small direct-product, $\mathbf{S}^{Srep}_L$ (Eq. \ref{Eq_Smolyak}).

\begin{equation}
\label{Eq_DP}
    \mathbf{S}^{DP}_{\bm{\ell}}=\mathbf{S}_{\ell_1}^1 \otimes \mathbf{S}_{\ell_2}^2 \otimes \cdots \otimes\mathbf{S}_{\ell_f}^f
\end{equation}

\begin{equation}
\label{Eq_Smolyak}
	\mathbf{S}^{Srep}_L=   \sum_{\mathcal{R}(\bm{\ell})} D_{\bm{\ell}} \cdot \mathbf{S}^{DP}_{\bm{\ell}} 
\end{equation}

In the previous equations, $f$, is the number of degrees of freedom (9 in the present study). $\mathcal{R}(\bm{\ell})$ is the constraint on the $\ell_i$ that enables control of the number of small direct products in the Smolyak expansion (Eq. \ref{Eq_Smolyak}). Usually, the constraint is defined as $L-f+1 \leq  \lvert \bm{\ell} \lvert \leq L$ where $L$ is the Smolyak parameter and $\lvert \bm{\ell} \lvert=\sum^f_{i=1} \ell_i$. With this constraint, the $D_{\bm{\ell}}$ coefficients are $(-1)^{L-\lvert \bm{\ell} \lvert} C_{f-1}^{L-\lvert \bm{\ell} \lvert}$ and $C_i^j$ are binomial coefficients.
Furthermore, in Eqs  \ref{Eq_DP},\ref{Eq_Smolyak}, the $\mathbf{S}$ represents, either a basis set,  $\mathbf{B}$, or a grid,  $\mathbf{G}$ and two $L$ parameters and control the basis set ($L_B$) and grid ($L_G$) sizes. The relation between, the $i^{th}$ primitive basis set ($\mathbf{B}_{\ell_i}^i$) or grid ($\mathbf{G}_{\ell_i}^i$) size and $\ell_i$ is given by: $nb_{\ell_i}^i = nq_{\ell_i}^i = A_i + B_i \cdot \ell_i$, where $A_i$ and $B_i$ are parameters. In the present study, the basis set and grid parameters,$A_i$ and $B_i$,  are all identical and equal to $1$ and $2$, respectively.

For the CHFClBr, the nine primitive basis sets are the normalized Harmonic Oscillator basis sets where the $j^{th}$ basis function associated to the the $i^{th}$ normal mode, $Q_i$, is:

\begin{equation}
\label{Eq_HO}
	b_j^i \left( Q_i \right) =  N_j \cdot H_{j-1} \left( Q_i \right)  \cdot Exp\left[ -\frac{1}{2}{Q_i}^2 \right]
\end{equation}
where, $N_j$ is a normalization coefficient, $H_{j-1}$ is the Hermite polynomial of degree $(j-1)$.

With these parameters, for $L_B=7$ and $L_G=9$, the basis set and the grid sizes are $224143$ and $14666470$, respectively. Therefore a direct diagonalization cannot be used, so the block Davidson iterative approach has been used. With these parameters, the convergence for the fundamental transitions (except for the one associated to the CH stretch) is better than $0.01$ \icm. 

\bibliography{refs}

\clearpage
\section{Supporting Information}

\subsection{Quartic Force Fields for water (unbound, bound)}\label{app:h2o_qff}

\textbf{unbound QFF}
\begin{lstlisting}
## QFF model ##
OP_DEFINE-SECTION
title
Water, 3D, QFF model
end-title
end-op_define-section

PARAMETER-SECTION

# Harmonic frequencies
w1  = 1611.108630 ,cm-1
w2  = 3826.060680 ,cm-1
w3  = 3930.213900 ,cm-1

# Cubic force constants
f111  = 282.030890 ,cm-1
f211  = -1059.682500 ,cm-1
f221  = -190.952160 ,cm-1
f222  = 1768.937060 ,cm-1
f311  = -0.544200 ,cm-1
f321  = 0.841260 ,cm-1
f322  = -2.302200 ,cm-1
f331  = -770.402580 ,cm-1
f332  = 5317.026990 ,cm-1
f333  = 2.210830 ,cm-1

# Quartic force constants
f1111  = -36.542630 ,cm-1
f2111  = 614.045640 ,cm-1
f2211  = -1860.612300 ,cm-1
f2221  = -239.329280 ,cm-1
f2222  = 733.774450 ,cm-1
f3111  = 0.132800 ,cm-1
f3211  = -8.498160 ,cm-1
f3221  = -0.271200 ,cm-1
f3222  = -12.945720 ,cm-1
f3311  = -2211.456840 ,cm-1
f3321  = -1393.687680 ,cm-1
f3322  = 4428.849720 ,cm-1
f3331  = -0.983480 ,cm-1
f3332  = -8.117960 ,cm-1
f3333  = 748.168570 ,cm-1

end-parameter-section

HAMILTONIAN-SECTION
    modes       |  v1 | v2 | v3
# KEO terms
1.0*w1    |1  KE
1.0*w2    |2  KE
1.0*w3    |3  KE

# QFF harmonic terms
0.5*w1     |1 q^2
0.5*w2     |2 q^2
0.5*w3     |3 q^2

# Cubic potential energy terms
f111/6.0      |1 q^3
f211/6.0      |1 q^2 |2 q
f221/6.0      |1 q |2 q^2
f222/6.0      |2 q^3
f311/6.0      |1 q^2 |3 q
f321/6.0      |1 q |2 q |3 q
f322/6.0      |2 q^2 |3 q
f331/6.0      |1 q |3 q^2
f332/6.0      |2 q |3 q^2
f333/6.0      |3 q^3

# Quartic potential energy terms
f1111/24.0      |1 q^4
f2111/24.0      |1 q^3 |2 q
f2211/24.0      |1 q^2 |2 q^2
f2221/24.0      |1 q |2 q^3
f2222/24.0      |2 q^4
f3111/24.0      |1 q^3 |3 q
f3211/24.0      |1 q^2 |2 q |3 q
f3221/24.0      |1 q |2 q^2 |3 q
f3222/24.0      |2 q^3 |3 q
f3311/24.0      |1 q^2 |3 q^2
f3321/24.0      |1 q |2 q |3 q^2
f3322/24.0      |2 q^2 |3 q^2
f3331/24.0      |1 q |3 q^3
f3332/24.0      |2 q |3 q^3
f3333/24.0      |3 q^4

end-hamiltonian-section

HAMILTONIAN-SECTION_v1
    modes       |  v1 | v2 | v3
1.0*w1    |1  KE
0.5*w1    |1  q^2
f111/6.0      |1 q^3
f1111/24.0      |1 q^4
end-hamiltonian-section

HAMILTONIAN-SECTION_v2
    modes       |  v1 | v2 | v3
1.0*w2    |2  KE
0.5*w2    |2  q^2
f222/6.0      |2 q^3
f2222/24.0      |2 q^4
end-hamiltonian-section

HAMILTONIAN-SECTION_v3
    modes       |  v1 | v2 | v3
1.0*w3    |3  KE
0.5*w3    |3  q^2
f333/6.0      |3 q^3
f3333/24.0      |3 q^4
end-hamiltonian-section

end-operator

\end{lstlisting}
\textbf{bound QFF}
\begin{lstlisting}
## bQFF model ##

OP_DEFINE-SECTION
title
Water, 3D, bQFF model
end-title
end-op_define-section

LABELS-SECTION
cq = Cross_bf[5.5,8]
q1C = Cubic_bf[8.39,24]
q1Q = Quartic_bf[8.39,24]
end-labels-section

PARAMETER-SECTION
#Same as QFF
end-parameter-section

HAMILTONIAN-SECTION
    modes       |  v1 | v2 | v3
# KEO terms
1.0*w1    |1  KE
1.0*w2    |2  KE
1.0*w3    |3  KE

# QFF harmonic terms
0.5*w1     |1 q^2
0.5*w2     |2 q^2
0.5*w3     |3 q^2

# Cubic potential energy terms
f111/6.0      |1 q1C
f211/6.0      |1 cq^2 |2 cq
f221/6.0      |1 cq |2 cq^2
f222/6.0      |2 q^3
f311/6.0      |1 cq^2 |3 cq
f321/6.0      |1 cq |2 cq |3 cq
f322/6.0      |2 cq^2 |3 cq
f331/6.0      |1 cq |3 cq^2
f332/6.0      |2 cq |3 cq^2
f333/6.0      |3 q^3

# Quartic potential energy terms
f1111/24.0      |1 q1Q
f2111/24.0      |1 cq^3 |2 cq
f2211/24.0      |1 cq^2 |2 cq^2
f2221/24.0      |1 cq |2 cq^3
f2222/24.0      |2 q^4
f3111/24.0      |1 cq^3 |3 cq
f3211/24.0      |1 cq^2 |2 cq |3 cq
f3221/24.0      |1 cq |2 cq^2 |3 cq
f3222/24.0      |2 cq^3 |3 cq
f3311/24.0      |1 cq^2 |3 cq^2
f3321/24.0      |1 cq |2 cq |3 cq^2
f3322/24.0      |2 cq^2 |3 cq^2
f3331/24.0      |1 cq |3 cq^3
f3332/24.0      |2 cq |3 cq^3
f3333/24.0      |3 q^4

end-hamiltonian-section

HAMILTONIAN-SECTION_v1
    modes       |  v1 | v2 | v3
1.0*w1    |1  KE
0.5*w1    |1  q^2
f111/6.0      |1 q^3
f1111/24.0      |1 q^4
end-hamiltonian-section

HAMILTONIAN-SECTION_v2
    modes       |  v1 | v2 | v3
1.0*w2    |2  KE
0.5*w2    |2  q^2
f222/6.0      |2 q^3
f2222/24.0      |2 q^4
end-hamiltonian-section

HAMILTONIAN-SECTION_v3
    modes       |  v1 | v2 | v3
1.0*w3    |3  KE
0.5*w3    |3  q^2
f333/6.0      |3 q^3
f3333/24.0      |3 q^4
end-hamiltonian-section

end-operator

\end{lstlisting}

\clearpage
\subsection{MCTDH input file for the GS-Water}
\begin{lstlisting}
### QFF model ##

RUN-SECTION
 name = GS
 title=H2O QFF model
 relaxation=0
 tfinal = 100.0  tout = all tpsi=1.0
 rlxunit=cm-1
 steps gridpop
end-run-section

OPERATOR-SECTION
  opname = mctdhbQFF
end-operator-section

SPF-BASIS-SECTION
 v1  =  15
 v2  =  15
 v3  =  15
end-spf-basis-section

PRIMITIVE-BASIS-SECTION
#Label  DVR  N   Parameters
  v1   HO    23   0.0     1.0     1.0
  v2   HO    23   0.0     1.0     1.0
  v3   HO    23   0.0     1.0     1.0
end-primitive-basis-section

INTEGRATOR-SECTION
  CMF     = 0.20, 1.0d-3
  RK8/spf = 1.0d-9
  rrDAV/A = 100, 1.0d-10
  eps_inv = 1.d-10
  natorb
end-integrator-section

INIT_WF-SECTION
 BUILD
  v1  eigenf v1 pop=1
  v2  eigenf v2 pop=1
  v3  eigenf v3 pop=1
 END-BUILD
end-init_WF-section
END-INPUT
\end{lstlisting}
\subsection{MCTDH input file for getting the fundamentals}
A lock relaxation (or propagation) is performed on the OPERATED GS which is read from a file as:
\begin{lstlisting}
INIT_WF-SECTION
 file=GS/restart
 operate=OP
end-init_WF-section
\end{lstlisting}
The initial wavepacket is obtained by generating nodal structure to the ground state wavefunction along the mode of interest. 
\begin{lstlisting}
HAMILTONIAN-SECTION_OP
    modes       |  v1 | v2 | v3 | v4 | v5 | v6 | v7 | v8 | v9
1.0    |1  q
end-hamiltonian-section
\end{lstlisting}
\subsection{Input for the computation of QFF using Gaussian 16}\label{app:g16}


\texttt{\#p mp2/aug-cc-pVTZ freq=(Anharmonic, NoRaman, HPmodes, ReadAnharm) guess=read geom=check nosymm
\\
CHFClBr VPT2
\\
0 1
\\
Print=NMOrder=AscNoIrrep}

\clearpage

 \section{Effect of Q$_\text{max}$ and n on the vibrational modes}
 While, the harmonic eigenfunctions can be used to predict the extent to which the correction perturbs each eigenstate, in this section, we investigate the effect of varying the parameters of the correction function applied to the cross terms ($n_2$ and $Q_\text{max}$) to check the convergence of the eigenstates. As a rule of thumb, for a specific $n_2$, if $Q_\text{max}$ is the highest value that eliminates all the holes, then values around 0.9$Q_\text{max}$ can be used. The lower $Q_\text{max}$, the less probable that the PES starts to decrease (before the correction), but it results in blue shifts of the eigenstates. 
 \\
 From Table 1, one can obverse that all reported eigenstates (ZPE, fundamentals, and combination bands) shows a clear convergence with increasing the value of $Q_\text{max}$. Some eigenstates converge faster than others depending on the extent of their anharmonicity. 
 \begin{table}[h]
\centering
\begin{tabular}{c c c c c c c c }
\hline
State/Q$_\text{max}$ &  2.5 & 3 & 3.5 & 4 & 4.5 & 5 & 5.5 \\ \hline
ZPE & 4641.5 & 4635.1 & 4632.8 & 4632.0 & 4631.7 & 4631.6 & 4631.5 \\ 
$\nu_1$ & 1555.7 & 1546.3 & 1541.9 & 1540.0 & 1539.2 & 1538.9 & 1538.7 \\ 
$\nu_2$ & 3729.5 & 3711.4 & 3704.2 & 3701.7 & 3700.8 & 3700.5 & 3700.4 \\ 
$\nu_3$ & 3863.9 & 3818.2 & 3797.5 & 3789.2 & 3786.0 & 3784.7 & 3784.2 \\ 
$\nu_1$ + $\nu_2$ & 5250.2 & 5220.1 & 5207.6 & 5202.6 & 5200.6 & 5199.8 & 5199.4 \\ 
$\nu_1$ + $\nu_3$ & 5383.4 & 5312.8 & 5278.9 & 5264.6 & 5258.8 & 5256.4 & 5255.4 \\ 
\hline
\end{tabular}
\caption{Vibrational modes of water at different values of Q$_\text{max}$ ($\mathbf{n_2}$=8)}
\end{table} 
\\
If necessary, one can use smaller values of $n_2$. The effect of this is that the region under $Q_\text{max}$ is now relatively more perturbed, compared to the higher values of $n_2$ as shown in Table 2. Moreover, it implies that the correction is slightly lower to eliminate the cross terms. Using higher values of $n_2$ perform the opposite. From our studied systems, fixing $n_2=8$ and varying $Q_\text{max}$ (depending on the system) is the most reliable and intuitive procedure.
\begin{table}[h]
\centering
\begin{tabular}{c c c c c c c c }
\hline
State/n &  2 & 4 & 6 & 8 & 10 & 12 & 18 \\ \hline
ZPE & 4655.4 & 4637.1 & 4633.0 & 4632.0 & 4631.7 & 4631.6 & 4631.5 \\ 
$\nu_1$ & 1566.8 & 1547.7 & 1541.9 & 1540.0 & 1539.3 & 1539.0 & 1538.7 \\
$\nu_2$ & 3734.1 & 3711.9 & 3704.1 & 3701.7 & 3700.9 & 3700.6 & 3700.3 \\ 
$\nu_3$ & 3910.1 & 3824.7 & 3797.7 & 3789.2 & 3786.2 & 3784.9 & 3784.0 \\ 
$\nu_1$ + $\nu_2$ & 5281.1 & 5225.5 & 5208.1 & 5202.6 & 5200.7 & 5200.0 & 5199.4 \\ 
$\nu_1$ + $\nu_3$ & 5451.0 & 5322.2 & 5279.0 & 5264.6 & 5259.2 & 5257.0 & 5255.1 \\ 
\hline
\end{tabular}
\caption{Vibrational modes of water at different values of $\mathbf{n_2}$ (Q$_\text{max}$=4)}
\end{table}
\newpage
\section{Comparision of QFF vs. bQFF eigenstates}
In the next 3 tables, we evaluate some eigenstates of QFF and compare it to bQFF results. Since these QFFs exhibit holes, the eigenstates could not ne straightforwardly  computed. The case of water was more prolematic and we had to keep decreasing the number of SPFs until no wave function leakage is observed. To maintain a consistent comparison, we re-performed the bQFF calculations with the same parameters. It can be observed in Table 3 that all the eigenstates are similar to the QFF results with differences less than 1 cm$^{-1}$, confirming the preservation of PES topology in the safe region. 
\begin{table}[h]
\centering
\begin{tabular}{c c c c c }
\hline
State  & VPT2  & QFF  & bQFF  \\
\hline
ZPE  & 4614.235 & 4631.507 & 4631.549 \\
$\nu_1$  & 1561.220 & 1538.609 & 1538.747 \\
$\nu_2$  & 3664.675 & 3700.299 & 3700.410 \\
$\nu_3$  & 3755.805 & 3784.081 & 3784.553 \\
$\nu_1$ + $\nu_2$  & 5215.461 & 5199.873 & 5200.125 \\
$\nu_1$ + $\nu_3$  & 5301.738 & 5255.086 & 5255.980 \\
\hline
\end{tabular}
\caption{Vibrational Modes of Water}
\end{table}
\\
The CXFClBr systems were less problematic. By simply starting from a wavefunction which is the product of the 1D eigenstates, the relaxation job smoothly converges without undergoing any wavefunction leakage. Similarly to water, the eigenstates (Table 4 and 5) are reproduced with bQFF with difference of ~1 cm$^{-1}$ at most are observed. 
\begin{table}[h]
\centering
\begin{tabular}{c c c c}
\hline
State & VPT2 & QFF  & bQFF   \\
\hline
ZPE & 4628.330 & 4627.164 & 4627.474 \\
$\nu_1$ & 228.476 & 228.349 & 228.369 \\
$\nu_2$ & 317.555 & 317.308 & 317.349 \\
$\nu_3$ &  427.916 & 427.717 & 427.774 \\
$\nu_4$ & 678.278 & 677.799 & 677.951 \\
$\nu_5$ & 794.246 & 794.530 & 794.717 \\
$\nu_6$ & 1066.989 & 1068.274 & 1068.422 \\
$\nu_7$ &  1226.018 & 1211.272 & 1212.455 \\
$\nu_8$ &  1309.824 & 1303.702 & 1304.720 \\
$\nu_9$ &  3059.131 & - & 3061 $\pm1$ \\
\hline
\end{tabular}
\caption{Vibrational Modes of CHFClBr}
\label{tab: CHFClBr modes}
\end{table}
\newpage
\begin{table}[h]
\centering
\begin{tabular}{c c c c}
\hline
State  & VPT2 & QFF  & bQFF   \\
\hline
ZPE &  3860.463 & 3860.118 & 3860.264 \\
$\nu_1$ &  227.794 & 227.693 & 227.709\\
$\nu_2$ &  317.054 & 316.870 & 316.905\\
$\nu_3$ &  425.831 & 425.640& 425.694\\
$\nu_4$ &  638.824 & 638.323& 638.445\\
$\nu_5$ &  756.198 & 756.338& 756.464\\
$\nu_6$ &  929.038 & 927.663 & 928.237\\
$\nu_7$ &  973.428 & 972.527 & 972.827\\
$\nu_8$ &  1072.615 & 1072.731 & 1072.948\\
$\nu_9$ &  2286.792 & - & 2288 $\pm1$\\
\hline
\end{tabular}
\caption{Vibrational Modes of CDFClBr}
\label{tab:CDFClBr modes}
\end{table}

\end{document}